# Peierls distortion, magnetism, and high hardness of manganese tetraboride

Huiyang Gou,[1,2] Alexander A. Tsirlin,[3*] Elena Bykova,[1,2] Artem M. Abakumov,[4] Gustaaf Van Tendeloo,[4] Asta Richter,[5] Sergey V. Ovsyannikov,[1] Alexander V. Kurnosov,[1] Dmytro M. Trots,[1] Zuzana Konôpková,[6] Hans-Peter Liermann,[6] Leonid Dubrovinsky,[1] Natalia Dubrovinskaia[2*]

[1] *Bayerisches Geoinstitut, Universität Bayreuth, D-95440 Bayreuth, Germany*

[2] *Material Physics and Technology at Extreme Conditions, Laboratory of Crystallography, University of Bayreuth, D-95440 Bayreuth, Germany*

[3] *National Institute of Chemical Physics and Biophysics, Akadeemia tee 23, E-12618 Tallinn, Estonia*

[4] *Electron Microscopy for Materials Research (EMAT), University of Antwerp, Groenenborgerlaan 171, B-2020 Antwerp, Belgium*

[5] *Technische Hochschule Wildau, Bahnhofstrasse 1, D-15745 Wildau, Germany*

[6] *DESY Photon Science, Deutsches Electronen Synchrtron, Notkestrasse 85, D-22607 Hamburg, Germany*

We report crystal structure, electronic structure, and magnetism of manganese tetraboride, $MnB_4$, synthesized under high-pressure high-temperature conditions. In contrast to superconducting $FeB_4$ and metallic $CrB_4$, which are both orthorhombic, $MnB_4$ features a monoclinic crystal structure. Its lower symmetry originates from a Peierls distortion of the Mn chains. This distortion nearly opens the gap at the Fermi level, but despite the strong dimerization and the proximity of $MnB_4$ to the insulating state, we find indications for a sizable paramagnetic effective moment of about 1.7 $\mu_B$/f.u., ferromagnetic spin correlations and, even more surprisingly, a prominent electronic contribution to the specific heat. However, no magnetic order has been observed in standard thermodynamic measurements down to 2 K. Altogether, this renders $MnB_4$ a structurally simple but microscopically enigmatic material; we argue that its properties may be influenced by electronic correlations.

PACS number(s): 61.50.-f, 62.20.-x, 75.20.En

## I. INTRODUCTION

The "electron-deficient" character of boron and its chemical activity lead to the formation of numerous boron-rich compounds of various structural complexity.[1-3] A plethora of interesting physical and chemical properties of boron-based solids, such as mechanical strength, high hardness, superconductivity, catalytic activity and thermoelectricity, keeps them in focus of modern experimental and theoretical research.[4-7]

The synthesis of diborides of 5$d$ noble metals, $OsB_2$ and $ReB_2$, was driven by expectations to obtain a new type of superhard materials[7-10] at ambient pressure. However, a careful analysis of the available data[11] and following investigations[12,13] did not confirm the proclaimed superhardness. Unexpected superhardness was found for iron tetraboride ($FeB_4$) synthesized at high pressures and temperatures[6], while other transition metal tetraborides (e.g. $CrB_4$ and $WB_4$)[14,15] are hard, but not superhard materials- their hardness is below 30 GPa in the asymptotic hardness region. Moreover, iron tetraboride was found to be superconducting,[6] thus possessing a combination of useful properties, which are desirable for a variety of engineering applications. This particular example motivated us for a further exploration of transition metal tetraborides, in particular, $MnB_4$. Its detailed structure investigation is still missing- the Inorganic Crystal Structure Database (ICSD) provides information about the monoclinic crystal structure of $MnB_4$ (space group $C2/m$), ICSD#15079, based on powder X-ray diffraction data of Andersson[16] and Andersson & Carlsson[17] obtained in late 1960s. So far $MnB_4$ has never been synthesized in a quantity sufficient for the investigation of its electronic and magnetic properties.

Here, we report the successful synthesis of single crystals of $MnB_4$ at high pressures and temperatures, solution and refinement of its crystal structure based on single-crystal X-ray diffraction, and results of investigations of the material's compressibility, hardness, magnetic properties, and electronic structure.

## II. MATERIALS AND METHODS

### A. Starting Materials and Synthesis

Polycrystalline $MnB_4$ samples were synthesized at high-pressure and high-temperature conditions in a piston-cylinder apparatus. Boron (Chempur Inc., 99.99% purity) and manganese (Alfa Aesar, 99.9% purity) powders were mixed in a stoichiometric (4:1) ratio. The mixture was loaded into a double capsule consisting of *h*-BN (inner) and Pt (outer) parts and then compressed to 3 GPa and heat treated either at 1080°C, 1350°C, or 1500°C. The duration of heating varied from 4 to 240 hours. The samples were abruptly quenched by switching off the furnace power. Pressure calibration was performed prior to the synthesis. It is based on the quartz-coesite and kyanite-sillimanite transitions, as well as on the melting point of diopside. The measured pressure is considered to be accurate within less than ± 5%. The temperature was measured with a Pt-Pt10%Rh thermocouple. Temperature gradients are estimated to be less than 25°C for the described experimental conditions.

Single-crystals of $MnB_4$ were synthesized at pressures of 10 and 12 GPa and a temperature of 1600°C (heating duration was 1 hour) in the Kawai-type multi-anvil apparatus[18] using 1000-ton (Hymag) and 1200-ton (Sumitomo) hydraulic presses and the 14/8 (octahedron edge length/cube truncation length) high-pressure assemblies. As starting materials we used a manganese rod (Goodfellow, 99.5% purity) and a boron powder (Chempur Inc., 99.99% purity) which were enclosed into a *h*-BN capsule. The pressure was calibrated based on the phase transitions of standard materials and the temperature was determined using a W3Re/W25Re thermocouple.

**B. Analytical techniques**

The morphology and chemical composition of the synthesized single crystals were studied by means of the scanning electron microscopy (SEM) (LEO-1530). Chemical purity of the samples was confirmed using wavelength dispersive X-ray (WDX) microprobe analysis (JEOL JXA-8200; focused beam; 12 keV and 15 nA or 15 keV and 12 nA). The LIFH and LDEB crystals were used to analyze Mn and B, respectively. Pure Mn and α-B or FeB were used as internal standards with ZAF correction.

**C. Single crystal X-ray diffraction**

A black lustrous thin plate of $MnB_4$ with a size of 0.05x0.04x0.01 $mm^3$ was used for the crystal structure investigation by means of single-crystal X-ray diffraction. X-ray diffraction data were collected at ambient temperature using a four-circle Oxford Diffraction Xcalibur diffractometer (λ = 0.7107Å) equipped with an Xcalibur Sapphire2 CCD detector. The intensities of the reflections were measured by step scans in omega-scanning with a narrow step width of 0.5°. The data collection and their further integration were performed with the CrysAlisPro software.[19] Absorption corrections were applied empirically by the Scale3 Abspack program implemented in CrysAlisPro. The scaling and absorption corrections were used due to the small size of the inspected crystal that makes precise face indexing difficult. The structure was solved by the direct method and refined by the full matrix least-squares in the anisotropic approximation for all atoms using SHELXTL software.[20] The X-ray experimental details and crystallographic characteristics of $MnB_4$ are presented in Table 1. The DIAMOND software[21] was used to create molecular graphics.

The crystallographic data of $MnB_4$ and further details of the crystal structure investigation have been deposited in the Inorganic Crystal Structure Database[22] and may be obtained free of charge from Fachinformationszentrum Karlsruhe, 76344 Eggenstein-Leopoldshafen, Germany (fax: (+49)7247-808-666; e-mail: crysdata@fiz-karlsruhe.de, http://www.fiz-karlsruhe.de/request_for_deposited_data.html) on quoting the deposition number CSD-426691.

**D. High-pressure powder X-ray diffraction**

For *in-situ* high-pressure X-ray diffraction studies we employed a piston-cylinder-type diamond anvil cell with a culet size of 350 μm and a rhenium gasket. A small sample (~ 20 μm in size) of a $MnB_4$ powder was loaded into a hole of ~ 150 μm in diameter drilled in the gasket pre-indented to ~ 50 μm. Using a gas-loading apparatus at BGI,[23] we loaded the pressure chamber with the sample along with neon as pressure-transmitting medium. The XRD experiments were carried out at the Extreme Conditions Beamline (ECB) P02.2 at PETRA III, DESY (Hamburg)[24]. The X-ray wavelength was $\lambda = 0.29135$ Å. The pressure was determined by the shift of the ruby luminescence line. The data were collected using a PerkinElmer XRD1621 detector and 2D X-ray images were integrated using the Fit2D program.[25]

**E. Transmission electron microscopy**

The sample for transmission electron microscopy (TEM) was prepared by crushing the material in an agate mortar under ethanol and depositing drops of the suspension on a holey carbon grid. The electron diffraction (ED) patterns and high resolution TEM (HRTEM) images have been acquired using a FEI Tecnai G2 microscope operated at 200 kV. Theoretical HRTEM images were calculated using the JEMS software.

**F. Hardness measurements**

Vickers hardness ($H_v$) was measured using a microhardness tester (M-400-G2, LECO Corporation) under loads of 0.5 kgf (4.9 N), 1 kgf (9.8 N) and 1.5 kgf (14.7 N).

Nanoindentation (NI) measurements were performed using the electrostatic transducer of the UBI 1 Hysitron triboscope with a pristine diamond 90° cube corner tip. We made several single (trapezoid) and multi-indentation measurements at 3 different areas on the sample with target loads 1.5/2.5/3.5/4.5/6 mN.

**G. Thermodynamic measurements**

The magnetic susceptibility was measured on small polycrystalline pieces of $MnB_4$ using the Quantum Design MPMS SQUID magnetometer. The data were collected at temperatures of 2–380 K in magnetic fields up to 5 T. The heat capacity was measured by a relaxation technique with the Quantum Design PPMS in the temperature range 1.8–200 K in fields of 0 and 5 T.

**H. Electronic structure calculations**

For electronic structure calculations, we used the full-potential local-orbital FPLO code[26] and the standard Perdew-Wang local density approximation (LDA)[27] for the exchange-correlation potential. The symmetry-irreducible part of the first Brillouin zone was sampled by a dense $k$ mesh of 518 points. The convergence with respect to the $k$ mesh was carefully checked.

**III. RESULTS**

**A. Crystal structure**

Based on powder X-ray diffraction data the crystal structure of $MnB_4$ was initially established as monoclinic[16,17] (space group $C2/m$). The reported unit cell parameters were $a = 5.5029(3)$, $b = 5.3669(3)$, $c = 2.9487(2)$ Å, $\beta = 122.710(5)°$ and the structure was described as a 3-dimensional boron network with Mn atoms inside the voids[17]. Each Mn atom is surrounded by 12 boron atoms and the distorted $MnB_{12}$ polyhedra pack in columns parallel to the $c$-direction (Fig. 1$a$) so that the metal atoms form one-dimensional chains with uniform Mn–Mn distances of 2.9487(2) Å. Every column of MnB12 polyhedra is shifted with respect to the four nearest ones over half of the *c* parameter. Andersson & Carlsson[17] described the structure of MnB$_4$ as highly similar to that of the orthorhombic CrB$_4$ (space group *Immm*)[28] with insignificant differences in the atomic arrangement.

Recent *ab initio* calculations[29] showed that the *M*B$_4$ (*M* = Cr, Mn, Fe, Tc, Ru) compounds are more stable if the structures have the symmetry described by the *Pnnm* space group. Indeed, investigation of the synthesized CrB$_4$ powder[29] by means of the electron and X-ray diffraction confirmed the existence of the orthorhombic (*Pnnm*) CrB$_4$ phase, whose structure was refined by Knappschneider *et al.*[14] based on single crystal X-ray diffraction data. Our recent studies[6] showed that FeB$_4$ has the same crystal structure as CrB$_4$. The unit cell contains three independent atoms: one *M*(1) atom in the (0, 0, 0) position at the center of inversion and two boron atoms, B(1) and B(2) in the 4*g* position. The major difference from the *Immm* structure (used by Andersson & Lundstroem[28] to describe the CrB$_4$ structure) is a distortion of the 3-dimensional boron network (Fig. 1*b*). Metal-metal distances in the *Pnnm* structures of CrB$_4$ and FeB$_4$ are 2.8659(1) and 2.9991(2) Å, respectively.

For MnB$_4$ we could expect the orthorhombic *Pnnm* crystal structure, but according to our findings, the β angle slightly differs from 90°. The distortion reduces the symmetry of the unit cell to monoclinic (*P*2$_1$/*n*) with *a* = 4.6306(3), *b* = 5.3657(4), *c* = 2.9482(2) Å and β = 90.307(6)°. Moreover in addition to the main reflections of the *Pnnm* subcell we have observed weak superstructure reflections corresponding to the **k** = (½,0,½) propagation vector. Using following transformation, **a**´ = **a** + **c**, **b**´ = –**b**, **c**´ = **a** – **c**, it was possible to index all reflections in the monoclinic unit cell (*P*2$_1$/*c*) with *a* = 5.4759(4), *b* = 5.3665(4), *c* = 5.5021(4) Å and β = 115.044(9)°. The unit cell of the MnB$_4$ structure contains five independent atoms (Mn(1) and B(1–4) atoms, see Table 2).

The obtained structure (Fig. 2) can be described in terms of the parent *Pnnm* cell plus a symmetry breaking structural distortion. The analysis of symmetry modes performed with the program AMPLIMODES[30,31] has shown that the *P*2$_1$/*c* distortion decomposes into two distortion modes of different symmetry corresponding to the irreducible representations (irreps) GM4+ and U1–.

The U1– irrep, associated with the **k**-vector (½,0,½) occurs as a primary mode for this distortion. It involves the displacements of Mn atoms along [1 0 1], thus resulting in two different Mn – Mn distances, namely 2.7004(6) and 3.1953(7) Å (Fig. 2*a*). This effect can be understood as a Peierls distortion of the Mn chains. In Fig. 3, we compare local density approximation (LDA) densities of states (DOS) calculated for the monoclinic $P2_1/c$ structure and for the orthorhombic *Pnnm* substructure, which is constructed as an "average" of the experimental $CrB_4$ and $FeB_4$ structures (averaged lattice parameters and atomic positions). In the orthorhombic structure, the Fermi level of $MnB_4$ would match the peak in the DOS, thus destabilizing the system. This effect is mitigated by a conventional Peierls distortion that splits the Mn chains with uniform Mn–Mn distances of about 2.93 Å into dimerized Mn chains with alternating Mn–Mn distances of 2.7004(6) and 3.1953(7) Å (as revealed by single-crystal X-ray diffraction). This way, the Fermi level falls into a dip of the DOS, which is unusual for transition-metal tetraborides. Indeed, both $CrB_4$ and $FeB_4$ remain orthorhombic and feature a relatively high DOS at the Fermi level, but remain stable with respect to the Peierls distortion.

Litterscheid *et al.*[32] recently reported in the conference abstracts about the growth of crystals of $MnB_4$ and its structure determination and refinement. However, neither synthesis was described, nor explicit structural information and details on the crystal structure investigation were given. The unit cell parameters were reported to be $a$ = 5.8982(2), $b$ = 5.3732(2), $c$ = 5.5112(2) Å and β = 122.633(3)°. They correspond to the choice of the non-standard unit cell with the space group $P2_1/n$, while the authors[32] provided the $P2_1/c$ space group.

The results of our TEM analysis are in agreement with the single-crystal XRD. Figure 4 shows the ED patterns of $MnB_4$. The patterns were indexed on a primitive monoclinic lattice with the cell parameters $a$ ≈ 5.5Å, $b$ ≈ 5.4 Å, $c$ ≈ 5.5 Å, β ≈ 115°, in agreement with the crystal structure determined from X-ray diffraction data. The [010] ED pattern (Fig. 4*d*) demonstrates apparent orthorhombic symmetry which results from a superposition of two mirror twinned variants of the monoclinic structure, shown in Fig. 4 (*e* and *f*). Taking into account twinning, the reflection conditions can be determined as $h0l$: $l$ = 2n and $0k0$: $k$ = 2n (Fig. 4*a-d*) that confirm the space group $P2_1/c$. The forbidden $0k0$, $k$ - odd reflections on the [001] and [100] ED patterns are caused by multiple diffraction as confirmed by the absence of these forbidden reflections in the [-101] ED pattern.

The [010] HRTEM image in Fig. 5 demonstrates that the $MnB_4$ crystal is almost free of extended defects. At these particular imaging conditions, the bright dots in the image correspond to projections of the Mn columns. The simulated HRTEM image, calculated with the crystal structure refined from single crystal X-ray diffraction data, is in excellent agreement with the experimental one. Figure 6 demonstrates a [010] HRTEM image of two twinned domains of the monoclinic $MnB_4$ structure. In spite of the coherent twin, no well-defined twin boundary separating the two domains is detected along this projection.

**B. Mechanical properties**

Figure 2*b* shows interatomic distances in $MnB_4$. The B–B distance of 1.703(6) Å is the shortest among the $MB_4$ ($M$ = Cr, Fe, Mn) compounds with similar crystal structures (see Table 3). According to Refs. 14 and 6, short B–B bonds are responsible for high hardness and low compressibility of $CrB_4$ and $FeB_4$, therefore we could expect similar properties in $MnB_4$.

The variations of the volume and lattice parameters of $MnB_4$ with pressure up to 25 GPa are presented in Fig. 7. The fit of the pressure-volume data with the third-order Birch-Murnaghan equation of state gave a bulk modulus of $K$ = 254(9) GPa and $K´$ =4.4 (Fig. 7*a*). The value of the bulk modulus is very close to that reported for $FeB_4$, 252(5) GPa.[6] Considerable anisotropy of the compressibility is also similar to that observed in $FeB_4$.[6] Along the *b* direction (Fig. 7*b*) the material is almost as incompressible as diamond[33]; this can be linked to the very short B-B bond (Figure 2*b*,Table 3) along the *b*-axis.

The Vickers hardness of the monoclinic $MnB_4$ was found to be 37.4 GPa at a load of 9.8 N and 34.6 GPa at 14.7 N; this is larger than that of the 5*d* transition metal borides, $WB_4$ (28.1 GPa[12] or 31.8 GPa[15] at 4.9 N), $ReB_2$ (18 GPa[13] at 9.8 N, 26.0-32.5 GPa[9] or 26.6 GPa[12] at 4.9 N), $OsB_2$ (19.6 GPa[8] or 16.8 GPa[12] at 4.9 N). Nanoindentation measurements resulted in the average hardness of 30.7 ± 2.3 GPa and the average indentation modulus of 415 ± 30 GPa. Thus $MnB_4$ is a fairly hard, but not superhard material. It is brittle, as indicated by the typical pop-ins and also cracks appearing sometimes after indentation and visible in the AFM images.

**C. Magnetic properties and electronic structure**

Magnetic susceptibility of $MnB_4$ reveals a weak ferromagnetic signal at low temperatures (Fig. 8). Above 150–200 K, $MnB_4$ shows the paramagnetic Curie-Weiss behaviour with the effective magnetic moment of 1.6–1.7 $\mu_B$ and the ferromagnetic Weiss temperature of $\theta \sim 90$ K according to

$$\chi = C/(T - \theta). \tag{1}$$

In Fig. 8 we show magnetic susceptibility data collected on two different samples which are both single-phase according to XRD and WDX. While the high-temperature regions match quite well, the behaviour at low temperatures is remarkably different and shows a variable magnitude of the ferromagnetic signal. Magnetization isotherms measured at 2 K further show a small, but variable net moment (Fig. 9). Therefore, we conclude that MnB$_4$ reveals ferromagnetic spin correlations evidenced by the positive $\theta$ value extracted from the robust high-temperature data. On the other hand, the low-temperature ferromagnetism of our samples (the net moment observed at low temperatures) appears to be extrinsic. Note also that no abrupt phase transition, such as ferromagnetic ordering, can be seen in the magnetization data.

Considering the LDA electronic structure of the stoichiometric monoclinic MnB$_4$ (Fig. 3, bottom), one would expect a weak paramagnetic or even a diamagnetic behaviour of this compound, because the Fermi level falls into a dip in the DOS formed upon the Peierls distortion, hence the number of states at the Fermi level is extremely low, only $N(E_F) \sim 0.08$ eV$^{-1}$/f.u., compared to $N(E_F) \sim 1.0$ eV$^{-1}$/f.u. in FeB$_4$. Surprisingly, our low-temperature heat-capacity measurements revealed a large electronic contribution to the specific heat. In the 15–30 K temperature range, the heat capacity can be fitted to the conventional expression for metals:

$$C_p(T) = \gamma T + \beta T^3, \tag{2}$$

where the first and second terms stand for the electronic and lattice contributions, respectively (Fig. 10). The fit yields $\gamma = 10.1$ mJ mol$^{-1}$ K$^{-2}$ and $\beta = 0.012$ mJ mol$^{-1}$ K$^{-4}$. Below 15 K, an additional contribution to the specific heat is clearly seen in Fig. 10. This contribution does not change in the applied field and may reflect non-magnetic impurity states leading to a series of Schottky anomalies. Its exact nature requires further investigation.

The $\beta$ and $\gamma$ parameters for MnB$_4$ are akin to those for FeB$_4$, where we previously reported $\gamma = 10.2$ mJ mol$^{-1}$ K$^{-2}$ and $\beta = 0.025$ mJ mol$^{-1}$ K$^{-4}$ (Ref. 6). Compared to superhard FeB$_4$, the $\beta$ value in MnB$_4$ is reduced by a factor of 2, which is well in line with our finding that MnB$_4$ is hard but not superhard. Its effective Debye temperature is $\theta_D \sim 540$ K, and the $\beta T^3$ behaviour of the lattice specific heat persists up to at least 30 K.

Regarding the electronic contribution to the specific heat, the $\gamma$ values of about 10 mJ mol$^{-1}$ K$^{-2}$ for MnB$_4$ and FeB$_4$ are remarkably similar. For a simple metal, they would imply a high density of states at the Fermi level, $N(E_F) \sim 4.3$ states eV$^{-1}$ f.u.$^{-1}$, which is four times higher than the LDA estimate for FeB$_4$ ($\sim 1.0$ states eV$^{-1}$ f.u.$^{-1}$, Ref. 5) and 50 times higher than the LDA estimate for MnB$_4$ ($\sim 0.08$ states eV$^{-1}$ f.u.$^{-1}$). Apparently, there is a strong renormalization of $\gamma$ in transition-metal tetraborides, yet in MnB$_4$ this effect is particularly strong. Possible reasons behind it will be discussed below.

The high value of $\gamma$ suggests that at least at low temperatures MnB$_4$ features a large number of charge carriers and should be metallic. While the small size of the available samples prevents us from performing the resistivity measurements, we note that already the large $\gamma$ value contradicts the simple scenario of a Peierls distortion that would drastically reduce the number of states at the Fermi level (Fig. 3, bottom). Moreover, ferromagnetic spin correlations can not be understood on the basis of LDA results, because the Peierls dimerization typically leads to a non-magnetic state. Indeed, spin-polarized LSDA calculations for MnB$_4$ converge to a non-magnetic solution, which contradicts the sizable effective moment and ferromagnetic spin correlations (positive $\theta$ value) observed in our magnetization measurements (Fig. 8).

The discrepancies between the non-magnetic, nearly insulating LDA scenario and the experimental ferromagnetic metallic behaviour can be ascribed to several effects. First, tiny deviations from the ideal MnB$_4$ stoichiometry may push the Fermi level out of the dip and increase the number of states at the Fermi level. However, this effect is by far insufficient to reproduce our results. A tentative modelling of the non-stoichiometric MnB$_4$ within the virtual crystal approximation (VCA) that basically changes the charge on the Mn site and shifts the Fermi level toward lower or higher energies, fails to account for ferromagnetic spin correlations: the system remains non-magnetic even at the 10% doping level, while the composition of our samples is established as stoichiometric MnB$_4$ with less than 1% uncertainty. A more plausible explanation would be an increased tendency to electron localization on the Mn sites. This tendency can be reproduced by the LSDA+$U$ method that adds a mean-field Hubbard-like energy term and mimics the effect of the on-site Coulomb repulsion $U$. Although originally designed for insulators, the LSDA+$U$ method can be also applied to metallic systems and provides a rough guess on the behaviour of correlated metals.[34]

Here, we used LSDA+$U$ with the on-site Coulomb repulsion $U$ = 3 eV and Hund's coupling $J$ = 0.5 eV, which were taken about twice lower than standard estimates for strongly correlated insulating Mn oxides ($U$ = 5−6 eV, $J$ = 1 eV, Refs. 35, 36). This way, we are able to stabilize a ferromagnetic solution with a small moment of about 0.6 $\mu_B$ on Mn atoms (Fig. 11). This moment is still much lower than the high-temperature paramagnetic effective moment of about 1.7 $\mu_B$. However, these two moments are not expected to match, because the LSDA+$U$ result pertains to the ordered moment at zero temperature, while the effective moment is the fluctuating moment at high temperatures. In fact, our calculated moment is in the same range as the ordered moment in Mn-based weak ferromagnets, such as MnSi: $\mu$ = 0.4 $\mu_B$ (Ref. 37). Moreover, we find a sizable density of states at the Fermi level, $N(E_F)$ ∼ 0.7 states eV$^{-1}$ f.u.$^{-1}$ (Fig. 11) that now approaches $N(E_F)$ ∼ 1.0 states eV$^{-1}$ f.u.$^{-1}$ for FeB$_4$ and better matches the experimental value of $\gamma$, although a large renormalization is still required.

**IV. DISCUSSION**

MnB$_4$ has its distinct position in the family of transition-metal tetraborides. Both CrB$_4$ and FeB$_4$ are orthorhombic and, in general, well described by standard LDA that accurately predicted the orthorhombic crystal structure of FeB$_4$ and even the superconductivity of this compound.[5,6] In MnB$_4$, the electron count is such that the Fermi level matches the maximum of the density of states. Then the orthorhombic structure becomes unstable and undergoes a monoclinic distortion. We ascribe this effect to a Peierls distortion, because in the monoclinic structure the Mn chains are dimerized, and the Fermi level falls into a dip in the density of states, which is strongly reminiscent of a band gap observed in other Peierls-distorted systems.[38,39] Therefore, MnB$_4$ could even be similar to narrow-gap intermetallic compounds, such as FeGa$_3$ (Ref. 40). An important difference though is that in those compounds the band gap would typically open because of the strong hybridization (formation of separated bonding and anti-bonding states) between the transition-metal and $p$-element orbitals. In MnB$_4$, the dip in the LDA DOS arises from the Mn−Mn interactions, while the mixing with the B states keeps the system metallic and provides a small yet non-zero number of states at the Fermi level.
Surprisingly, our experimental data are not consistent with this simple dimerization picture, because MnB$_4$ shows a large electronic contribution to the specific heat and a sizable high-temperature paramagnetic moment with clear signatures of ferromagnetic spin correlations.

Phenomenologically, MnB$_4$ is similar to Mn-based ferromagnets, such as MnSi (compare, for example, the high-temperature paramagnetic moments of ~1.7 μ$_B$ and 2.3 μ$_B$ (Ref. 37), respectively), with the only exception that MnB$_4$ does not show any clear signature of the long-range ferromagnetic order. We have shown that moderate electronic correlations may reconcile experimental observations with computational results and render MnB$_4$ ferromagnetic. However, the origin of these correlations is presently unclear, and the absence of the long-range magnetic ordering despite sizable ferromagnetic spin correlations remains an open problem as well.

## V. CONCLUSION

The high-pressure high-temperature synthesis technique enabled us to synthesize high-quality single crystals of manganese tetraboride, MnB$_4$. Single-crystal synchrotron X-ray diffraction data allowed the refinement of its crystal structure, which revealed dimerized Mn chains with alternating Mn–Mn distances, which were not identified in previous powder-XRD investigations of polycrystalline MnB$_4$ samples. We explained this phenomenon by a Peierls distortion, which reduces the symmetry of MnB$_4$ to monoclinic, compared to the orthorhombic symmetry of otherwise similar CrB$_4$ and FeB$_4$ structures. Mechanical property measurements revealed the high bulk modulus (254(9) GPa), strong anisotropy in compressibility (with the stiffness comparable to that of diamond, along the *b* axis), and very high hardness (35-37 GPa) approaching that of superhard materials. Our experimental studies provide previously unavailable data on magnetic properties of MnB$_4$. The latter, complemented with our theoretical consideration on the electronic properties of MnB$_4$, allowed us to conclude that the relatively simple crystal structure with a well-defined and well-understood Peierls distortion hosts remarkably complex and even enigmatic low-temperature physics. Current efforts in the high-pressure synthesis should eventually result in the preparation of larger samples that would facilitate further studies on the electronic structure and magnetism of this interesting material.

**ACKNOWLEDGMENTS**

H.G. gratefully acknowledges financial support of the Alexander von Humboldt Foundation. The work was supported by the German Research Foundation (DFG). N.D. thanks DFG for financial support through the Heisenberg Program and the DFG Project DU 954-8/1. A.T. was funded by the Mobilitas program of the ESF (grant no. MTT77). AT acknowledges fruitful discussions with Christoph Geibel and experimental support by Deepa Kasinathan. GVT acknowledges the European Research Council, ERC grant N°246791 – COUNTATOMS. Portions of this research were carried out at the light source PETRA III at DESY, a member of the Helmholtz Association (HGF).

Table 1. Details on the X-ray diffraction data collection and structure refinement of MnB$_4$.

| | |
|---|---|
| Empirical formula | MnB$_4$ |
| Formula weight (g/mol) | 98.18 |
| Temperature (K) | 296(2) |
| Wavelength (Å) | 0.7107 |
| Crystal system | Monoclinic |
| Space group | *P*2$_1$/*c* |
| *a* (Å) | 5.4759(4) |
| *b* (Å) | 5.3665(4) |
| *c* (Å) | 5.5021(4) |
| *β* (°) | 115.044(9) |
| *V* (Å$^3$) | 146.486(19) |
| Z | 4 |
| Calculated density (g/cm$^3$) | 4.452 |
| Linear absorption coefficient (mm$^{-1}$) | 8.319 |
| F(000) | 180 |
| Crystal size (mm$^3$) | 0.05 x 0.04 x 0.01 |
| Theta range for data collection (deg.) | 4.11 to 34.57 |
| Completeness to theta = 27.59° | 100.0 % |
| Index ranges | -8 < *h* < 8, |
| | -8 < *k* < 7, |
| | -8 < *l* < 8 |
| Reflections collected | 2122 |
| Independent reflections / $R_{int}$ | 593 / 0.0467 |
| Max. and min. transmission | 1.00000 and 0.78298 |
| Refinement method | Full matrix least squares on *F*$^2$ |
| Data / restraints / parameters | 593 / 0 / 34 |
| Goodness of fit on *F*$^2$ | 1.043 |
| Final *R* indices [*I* > 2σ(*I*)] | $R_1$ = 0.0376, $wR_2$ = 0.0731 |
| *R* indices (all data) | $R_1$ = 0.0652, $wR_2$ = 0.0813 |
| Largest diff. peak and hole (e / Å$^3$) | 0.728 and -0.911 |

Table 2. Atomic coordinates, positions and equivalent isotropic displacement parameters for MnB$_4$.

| Atom | Wykoff site | $x$ | $y$ | $z$ | $U_{eq}{}^a$, Å$^2$ |
|---|---|---|---|---|---|
| Mn(1) | 4$e$ | 0.26817(9) | 0.0011(2) | 0.273758) | 0.00465(15) |
| B(1) | 4$e$ | 0.3648(9) | 0.1859(8) | 0.6378(8) | 0.0072(4)$^b$ |
| B(2) | 4$e$ | 0.6699(8) | 0.1302(8) | 0.3238(8) | 0.0067(4)$^b$ |
| B(3) | 4$e$ | 0.8692(9) | 0.1822(8) | 0.1269(8) | 0.0072(4)$^b$ |
| B(4) | 4$e$ | 0.1639(8) | 0.1301(8) | 0.8405(8) | 0.0067(4)$^b$ |

$^a$ $U_{eq}$ is defined as one third of the trace of the orthogonalized $U^{ij}$ tensor.

$^b$ ADPs for B(1) and B(3) and for B(2) and B(4) have been fixed to be equal to each other.

Table 3. Bond lengths in $M$B$_4$ ($M$ = Mn, Cr, Fe) possessing similar structures.

| Metal boride | $M$–B distances, Å | B–B distances, Å | Reference |
|---|---|---|---|
| MnB$_4$ | 1.999(4)–2.310(4) | 1.703(6)–1.893(8) | This work |
| CrB$_4$ | 2.053(4) | 1.743(6) | 14 |
| | 2.153(4) | 1.835(4) | |
| | 2.178(3) | 1.868(6) | |
| | 2.261(3) | | |
| FeB$_4$ | 2.009(4) | 1.714(6) | 6 |
| | 2.109(4) | 1.8443(3) | |
| | 2.136(3) | 1.894(6) | |
| | 2.266(3) | | |

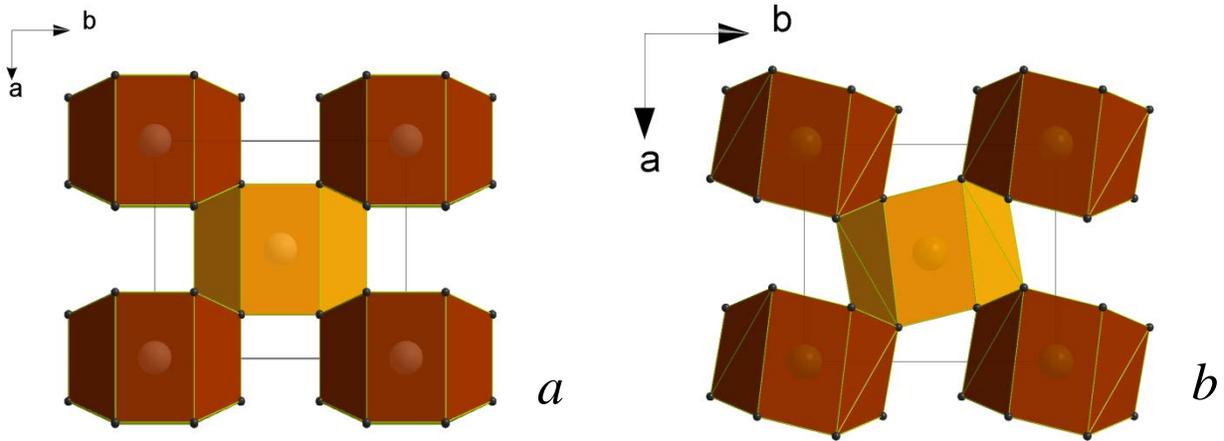

FIG. 1. (Color online) A comparison of the crystal structure of MnB$_4$ proposed by Andersson[16] (*a*), and that of FeB$_4$ (Ref.6) structure (*b*). In both cases *M*B12 polyhedra pack in columns, each one is shifted on a *c*/2 distance along the c-direction with respect to its four nearest neighbors (light and dark polyhedra), however a distortion of the 3-dimentional boron network is different.

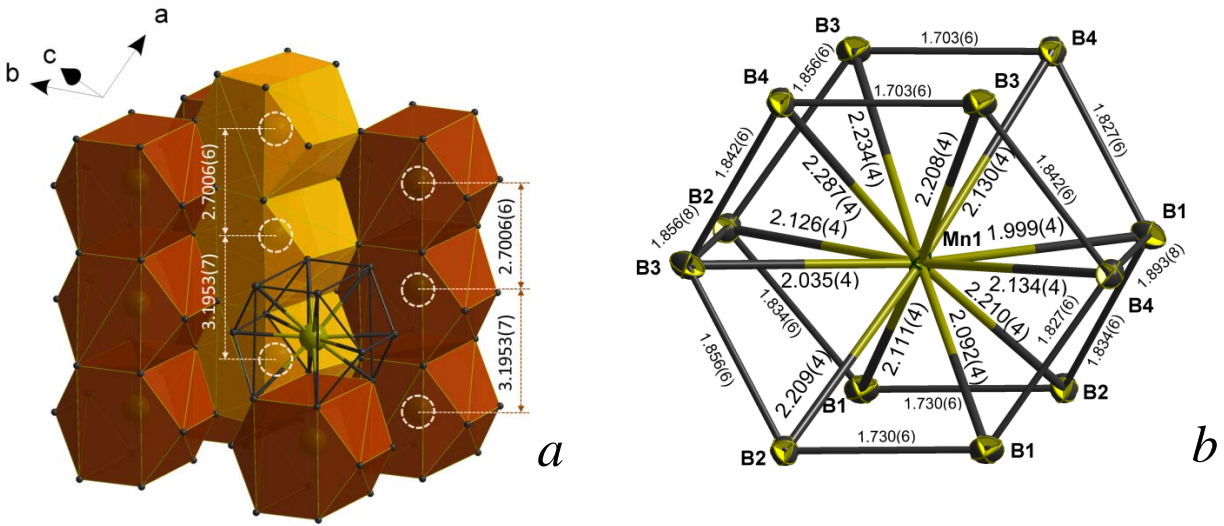

FIG. 2. (Color online) Structure of MnB$_4$. MnB12 polyhedra pack in columns along [1 0 1] direction with alternating Mn–Mn distances of 2.7006(6) and 3.1953(7) Å through the column (*a*). Interatomic distances (Å) in the MnB12 polyhedron (*b*).

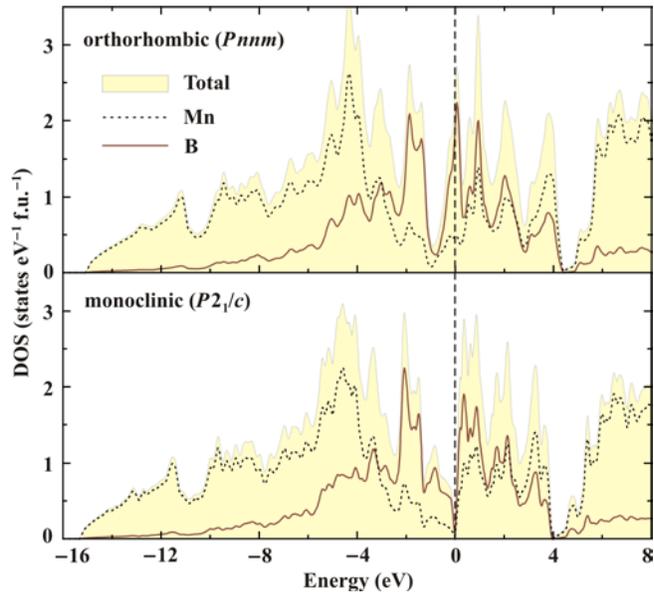

FIG. 3. (Color online) LDA DOS for MnB$_4$ in its fictitious FeB$_4$-like (orthorhombic, top panel) and real (monoclinic, bottom panel) structures. The monoclinic distortion shifts the Fermi level away from the DOS maximum and nearly opens a gap.

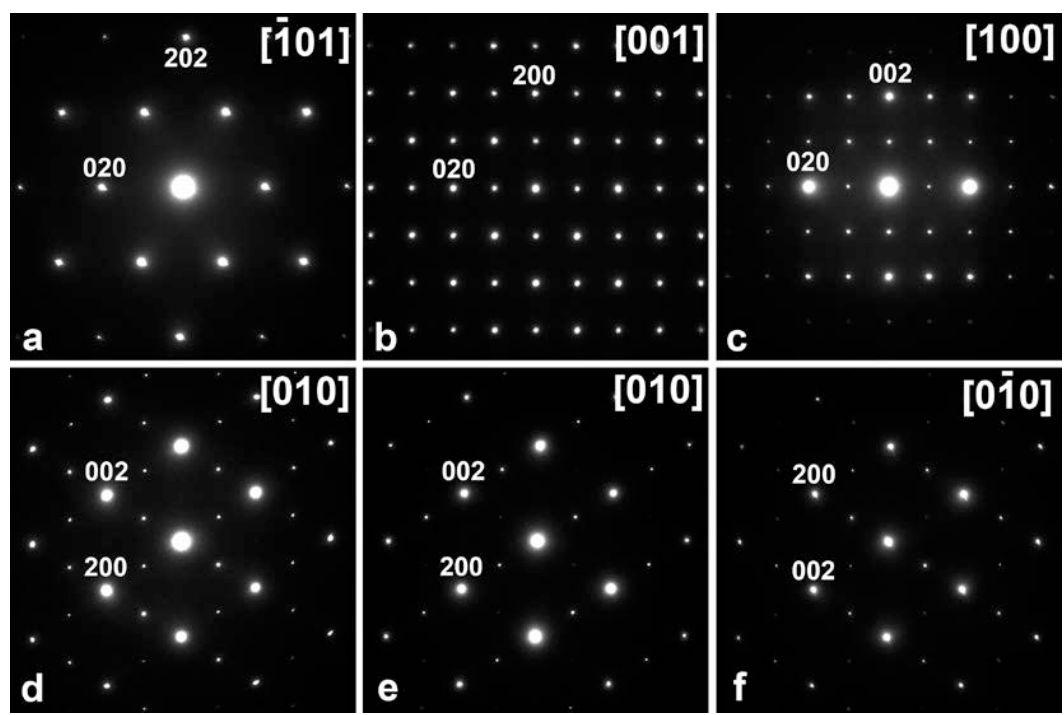

FIG. 4. Electron diffraction patterns of MnB$_4$. The [010] ED pattern (d) is a superposition of two twinned variants (e) and (f).

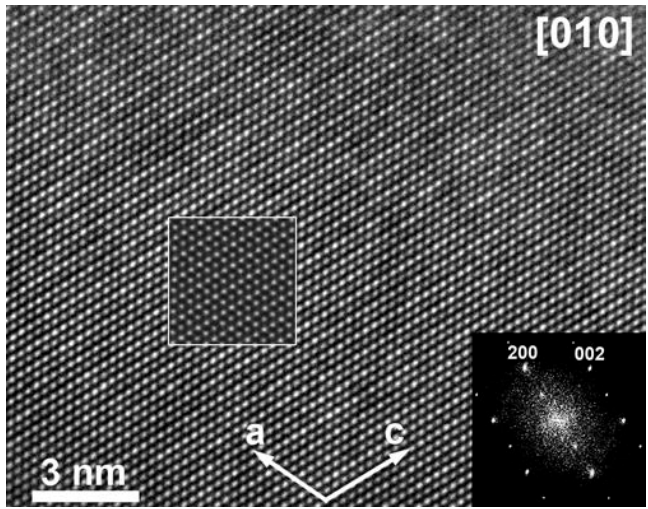

FIG. 5. [010] HRTEM image of a single domain of MnB$_4$ and its Fourier transform. The insert shows a calculated HRTEM image (defocus $f = 7$ nm, thickness $t = 4.8$ nm).

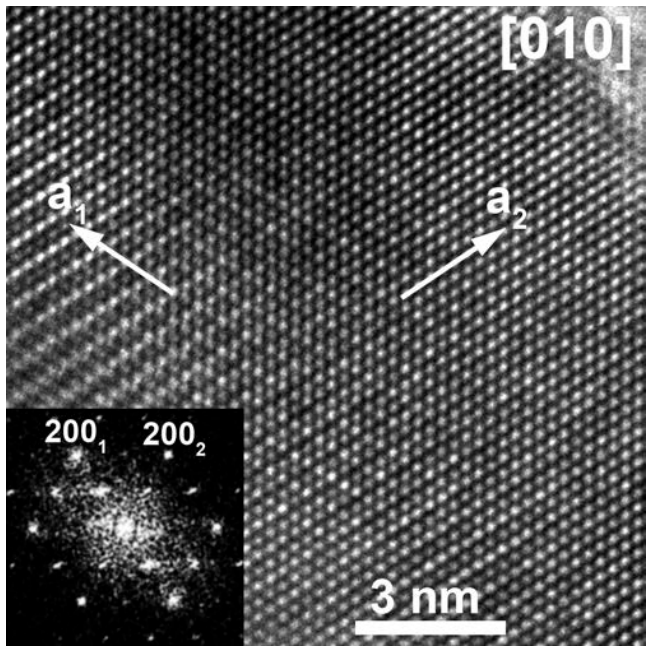

FIG. 6. [010] HRTEM image of two twinned domains of the MnB$_4$ structure (at the left and right side of the image, respectively) and corresponding Fourier transform showing two mirror-related orientations of the *a*-axis of the domains. No well-defined twin boundary is visible along this zone axis.

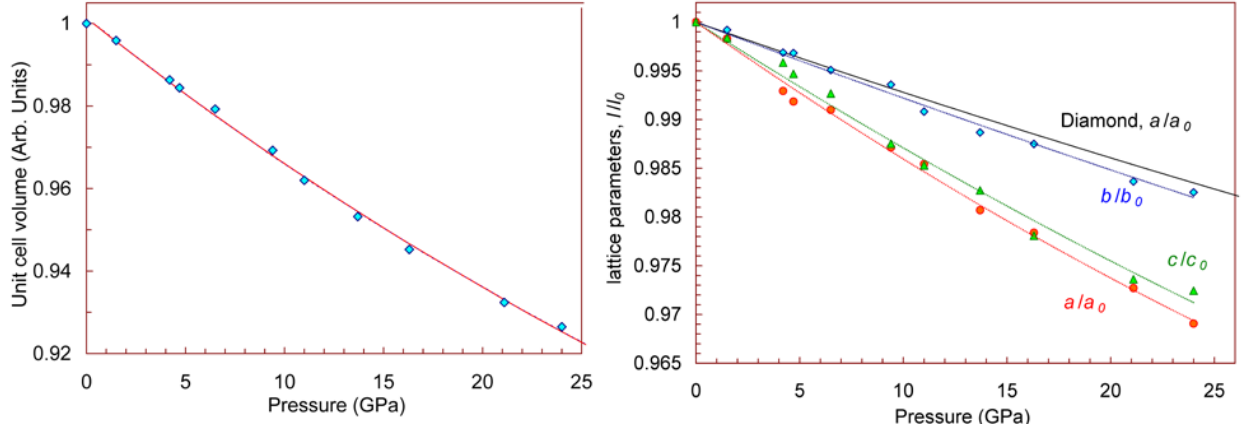

FIG. 7. (Color online) Compressibility of $MnB_4$. (*a*) The pressure dependence of the unit cell volume based on powder synchrotron X-ray diffraction data. Solid line corresponds to the fit of the pressure-volume data with the third-order Birch-Murnaghan equation of state, which gave the bulk modulus $K$ = 254(9) GPa and $K´$ =4.4. (*b*) The relative changes of the unit cell parameters as a function of pressure. The stiffness of the $MnB_4$ structure along the *b*-direction is almost the same as that of diamond (continues black line according to Ref. 33).

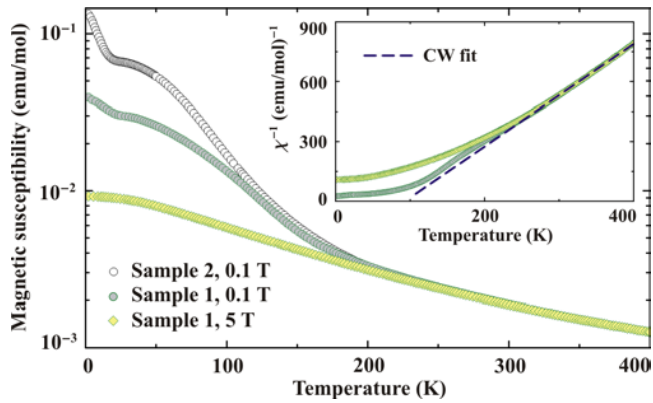

FIG. 8. (Color online) Magnetic susceptibility of MnB$_4$ measured on two different samples. At high temperatures, the susceptibility is nearly sample-independent and yields the Curie-Weiss (CW) parameters of $\mu_{eff}$ ~ 1.7 $\mu_B$ and $\theta$ ~ 90 K (see inset). At low temperatures, the susceptibility is strongly sample-dependent indicating a variable net moment, which is most likely extrinsic.

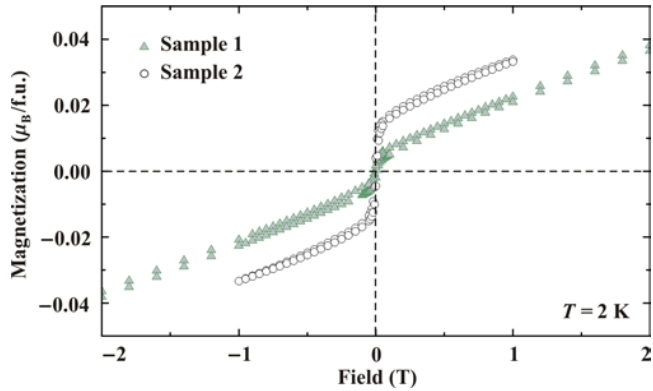

FIG. 9. (Color online) Magnetization curves of two MnB$_4$ samples measured at 2 K. Note the different net moments and the similar slope of the linear part.

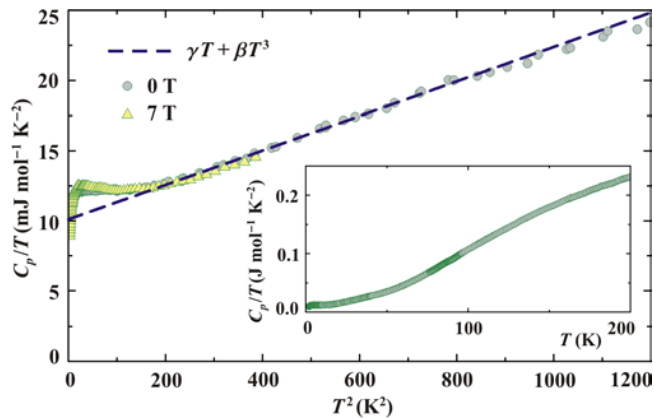

FIG. 10. (Color online) The specific heat of MnB$_4$ measured in the applied fields of 0 T (circles) and 7 T (triangles). The line shows the fit with Eq. (2). The inset displays the smooth temperature dependence of the specific heat in a broad temperature range up to 200 K.

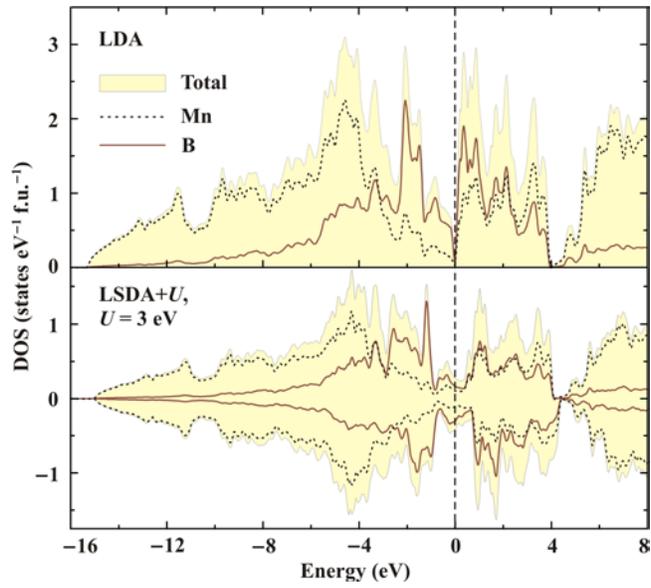

FIG. 11. (Color online) Electronic structure of monoclinic MnB$_4$ calculated within LDA (top panel) and LSDA+$U$ with $U = 3$ eV (bottom).